# Energy Efficient Homes: The Social and Spatial Patterns of Residential Energy Efficiency in England

Boyana Buyuklieva[1][1], Adam Dennett[2][1], Nick Bailey[3][2] and Jeremy Morley[4][3]

[1]Bartlett Centre for Advanced Spatial Analysis, University College London
[2]Department of Urban Studies, University of Glasgow
[3]Ordnance Survey Ltd

**Summary**

Poor energy efficiency of homes is a major problem with urgent environmental and social implications. Housing in the UK relies heavily on fossil fuels for energy supply and has some of the lowest energy efficiency in Europe. We explore spatial variations in energy efficiency across England using data from Energy Performance Certificates (EPCs), which cover approximately half of the residential stock (14M homes between 2008-22). We examine variations between authorities after accounting for the composition of the housing stock in terms of its 'fixed' characteristics of property type, building age and size. We explore variations in terms of geographical and social context (region, urban-rural and deprivation), which gives a picture of the scale of the challenge each faces. We also examine variations in relation to the more readily upgraded factors, such as glazing types, and in relation to local participation in improvement programmes which gives some insight into local actions or progress achieved.

1. **Introduction**

   1.1. **Motivation**

Improving domestic energy efficiency is vital if the UK Government is to meet its Net-Zero policy commitments by 2050 as approximately one-fifth of carbon emissions in the UK are linked to domestic energy usage (Woodland, 2021). Government modelling suggests emissions from buildings need to be virtually eliminated if Net Zero is to be achieved (Skidmore, 2023). Alongside the environmental motivation is a parallel social issue: recent increases in energy prices have pushed some households further into poverty. In 2022, the ONS estimated that the poorest 10% of households were spending more than half of their weekly expenditure on essentials such as electricity and gas (Woodward, 2021). In parallel, the political and policy landscape across the country varies considerably with different local authorities prioritising different needs and themes at different times (LGA, 2022; Simmie, 1994). Moving toward a sustainable and equitable domestic energy usage landscape will require a range of different strategies that take into account the heterogeneity of domestic housing stock and varied living conditions across places across the country.

   1.2. **Energy Performance Certificates**

Energy Performance Certificates (EPCs) were introduced in the UK in 2007 as a way to assess properties' energy efficiency. Energy efficiency is defined as the amount of energy required per square metre to achieve a given standard (i.e. heating the property to a specified regime, supplying hot water, and providing power for lighting, appliances, etc.). Ratings were designed to run from 1 to 100, with the latter indicating zero net energy consumption. Some dwellings now have ratings above 100 as they are net generators.

---

[1] boyana.buyuklieva@ucl.ac.uk
[2] a.dennett@ucl.ac.uk
[3] nick.bailey@glasgow.ac.uk
[4] jeremy.morley@os.uk

Properties for sale or for rent must have an EPC not more than ten years old. They are also required when accessing Net Zero-related government funding schemes. For example, EPCs necessary for private rentals by the Minimum Energy Efficiency Standards (MEES) regulations, and EPCs are needed for accessing programmes such as the Green Deal, which aim to help households make improvements to their homes.

### 1.3. Scope

Geographies of domestic energy efficiency are inevitable and need to be considered for the making of impactful policy. This paper extends previous modelling efforts (Bower, 2022; Wood, 2022) to ask where there is scope for affecting change, given the fixed characteristics of properties, popular interventions and local deprivation. It disentangles the various factors contributing to differences in energy efficiency levels across English local authorities using the Department for Levelling Up, Housing and Communities' (DLUHC) Energy Performance Certificate (EPC) dataset. After data processing to create a cross-sectional overview of EPCs, we control for some immutable - or at least slow-to-change - effects at property level to estimate the baseline differences between local authorities, controlling for differences in the basic mix or composition of stock. These more immutable property characteristics include age, size, and morphology - e.g. being a flat or detached house. We allow for variations in the mix of tenures since building standards and hence energy efficiency standards vary across these groups. We further use a residual analysis to highlight the resulting geography of energy efficiency challenges that authorities are facing and explore variations by location and social context.

## 2 Analysis

### 2.1. Data

EPCs are generated using manual inputs from surveys into regulated software programs that implement the latest versions of the Standard Assessment Procedure (SAP), the government's approved methodology of measuring energy performance for building regulation compliance (Stroma 2016). Assessments provide basic property characteristics (age, type, floor area) as well as tenure. Locational data is available from the location and includes local authority and region. There are two methodologies for the procedure: the full SAP for new dwelling including those that are produced through change of use; and reduced data SAP (RdSAP) for existing properties (Stroma 2016), with three versions of the calculation methodologies since 2007. Crawley et al. (2019, Figure. 1) suggest changes in the methodology to produce EPC scores have little impact on the distribution of energy efficiency ratings over time.

Two measures of social deprivation have been attached to properties at the Lower Layer Super Output Area (LSOA) level. These are: 1) the proportion of people Income Deprived on the 2019 English Indices of Multiple Deprivation and 2) the English Indices of Multiple Deprivation itself. We also create an urban-rural indicator derived from the property's location at the same level based on the Office for National Statistics' 2011 rural-urban classification.

### 2.2. Methodology

In our baseline model, we examine the energy efficiency of domestic properties as a function of a series of fixed effect characteristics. We then further explore the model residuals to measure whether properties in an authority are either over- or under-performing relative to these baseline features. Our model (1) – where all variables are categorical except for size – is as follows:

$$E.Efficiency = Intercept + \beta_1\, age + \beta_2\, type + \beta_3\, tenure + \beta_4\, log(size) + \beta_5\, localAuthority + \varepsilon \qquad (1)$$

We use the residuals of the base model (effectively variations within each local authority) to ask: how can we explain the prediction error after controlling for the characteristics of the local building stock? This time we examine how property-level alterations and area-level characteristics capture the divergence between expected and actual energy efficiency. To summarise these, we look at the proportion of the variance in residuals which is explained by each factor.

### 2.3. Descriptive Overview

The mean and median EPC rating is in band 'D' with a sharp rise in properties at the bottom of the 'E' grade at 39 points. This 'spike' in the distribution likely reflects the requirement for rental properties to achieve at least the 'E' letter score. Furthermore, older homes – houses built before the mid-1970s – perform less well. These are often owner-occupied, the majority of the recorded stock and most likely to score below 69 (band 'C'). In addition, we observe a small number of extreme EPCs with scores of '1'( ~ 0.1%, not shown). These tend to be marketed sales properties built pre-1900. As these are few but influential, they are excluded from modelling.

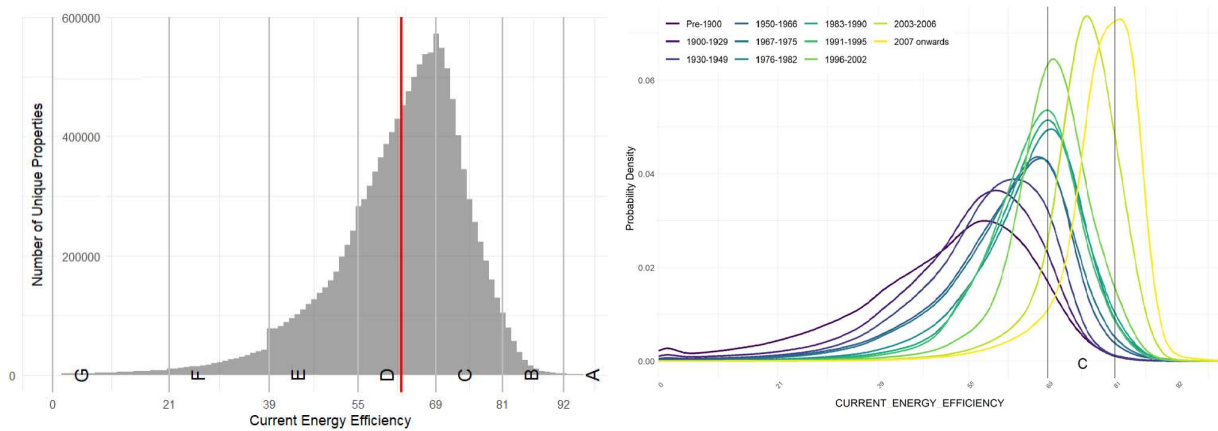

**Figure 1** Energy Performance Rating (left) relative to stock age (right)

### 2.4. Base Model

Fitting a standard linear model to our dataset, we can explain almost one third of the variance in energy performance by accounting for the age, type, tenure and local authority location. Age is the main influence, with properties showing substantial improvements over time - a reflection in large part of the success of regulation through building standards. More recent buildings perform on average much better, but interestingly, those properties built between 2007-2011 appear to perform better than the more modern baseline classification (2012 onwards).

**Table 1** Base Model (1) Summary

**Base Model:**
Construction Age + Dwelling Type + Tenure + log(Floor Area) + Local Authority (LA)

| | | adj.R2 | RSE | rsd.median | terms |
|---|---|---|---|---|---|
| | **Model Summary:** | 0.32 | 9.83 | 1.52 | 332 |

| | Term | estimate | statistic | p.value | std.error |
|---|---|---|---|---|---|
| 1 | **Intercept** | 81.5 | 1447.5 | 0 | 0.1 |
| | **Construction Age (β1)** | | | | |
| | Reference: Built from 2012 onwards (newest) | | | | |
| 2 | 2007-2011 | 0.9 | 27.5 | 0 | 0 |
| 3 | 2003-2006 | -1.4 | -58.6 | 0 | 0 |
| 4 | 1996-2002 | -5.8 | -246.1 | 0 | 0 |
| 5 | 1976-1995 | -9.4 | -440.4 | 0 | 0 |
| 6 | 1950-1975 | -13.4 | -635 | 0 | 0 |
| 7 | 1930-1949 | -16.3 | -744.5 | 0 | 0 |
| 8 | 1900-1929 | -19.3 | -890.1 | 0 | 0 |
| 9 | pre-1900 | -22.4 | -1008.1 | 0 | 0 |
| | **Dwelling Type (β2)** | | | | |
| | Reference: Flat/Maisonette (highest scoring EPC) | | | | |
| 10 | Terraced House | -1.3 | -148.1 | 0 | 0 |
| 11 | Semi-Detached | -3.5 | -347.4 | 0 | 0 |
| 12 | Detached | -5.9 | -451.7 | 0 | 0 |
| 13 | Bungalow | -5.7 | -481.6 | 0 | 0 |
| | **Tenure (β3)** | | | | |
| | Reference: Owner Occupied (majority group) | | | | |
| 14 | Privately Rented | 0.4 | 55.5 | 0 | 0 |
| 15 | Socially Rented | 4.5 | 561.6 | 0 | 0 |
| | **Floor Area (β4)** | | | | |
| 16 | Total Floor Area (log10) | -0.6 | -65.9 | 0 | 0 |

**Note:** Local authority fixed effects are shown in Table 2

Table 2 Summary of LA Fixed Effects Estimates by Region

**Local Authority (β5)**

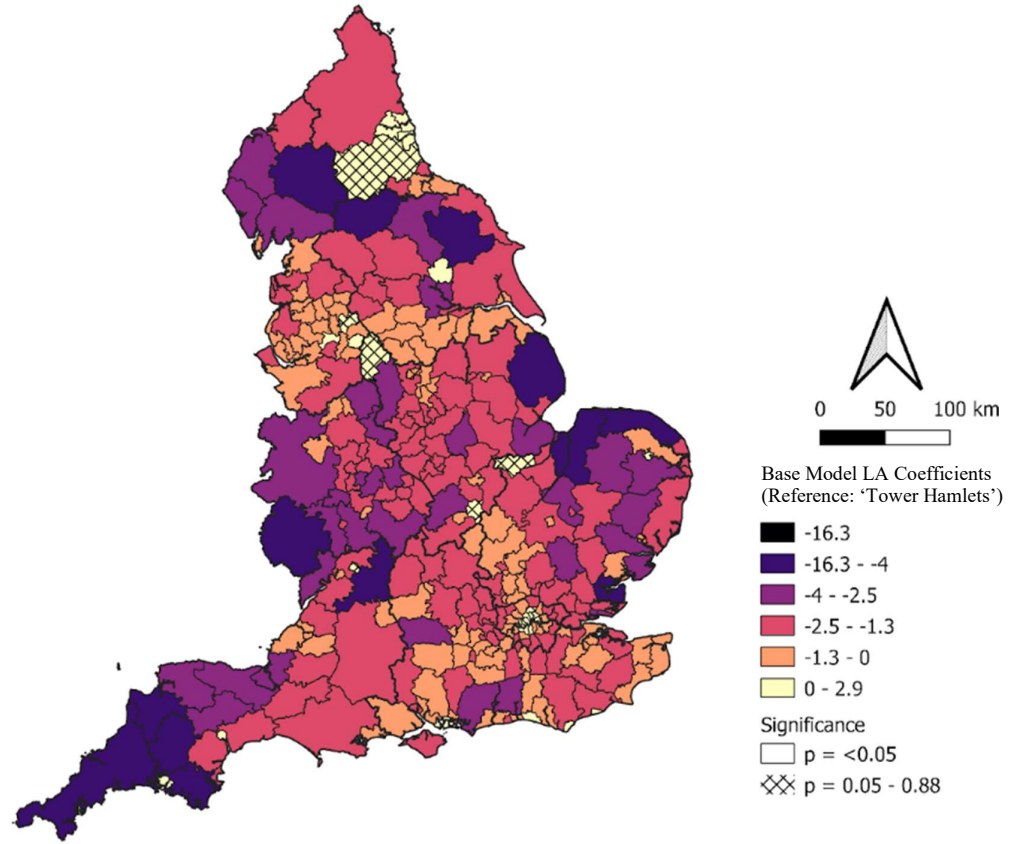

*See map for coefficient estimates.*
*Estimate summaries are provided below:*

| | Region | est.range | median | mean | SD | n (p < 0.05)/n |
|---|---|---|---|---|---|---|
| 17 | North East | 2.4 | -0.2 | -0.42 | 0.9 | 1/12 |
| 18 | London | 5.4 | -0.6 | -0.48 | 1.5 | 1/32 |
| 19 | North West | 7.2 | -1.0 | -1.31 | 1.3 | 1/39 |
| 20 | South East | 4.6 | -1.1 | -1.17 | 0.8 | 1/67 |
| 21 | Yorkshire and The Humber | 6.2 | -1.5 | -1.82 | 1.5 | 0/21 |
| 22 | East Midlands | 4.1 | -1.9 | -1.77 | 0.9 | 3/40 |
| 23 | East of England | 5.2 | -1.9 | -1.9 | 1.1 | 2/45 |
| 24 | West Midlands | 3.7 | -2.0 | -2.18 | 0.9 | 0/30 |
| 25 | South West | 16.7 | -2.1 | -2.74 | 3.1 | 1/30 |

After age, dwelling type is an important factor, with flats (with fewer external faces and hence lower heat loss) performing the best. The most inefficient homes tend to be larger detached houses, with bungalows and those with fewer external walls (semi-detached and terraced) performing much better. Tenure is also important, with those living in social housing (relative to owner occupation) faring considerably better in energy performance. Privately rented units also tend to be more energy efficient. However, there is a confounding effect of many private rental properties being flats. The most efficient homes tend to be smaller, newer and socially rented flats - especially those built after 2006. The worst performing stock is owner-occupied, then private and social rent. This may mirror how these tenures are regulated. Size remains important in these models even though the dependent is energy efficiency (i.e. energy use per square meter). Larger properties have lower energy efficiency.

## 2.5. Residual Analysis

**Table 3** Summary of Base Model Residual by Region

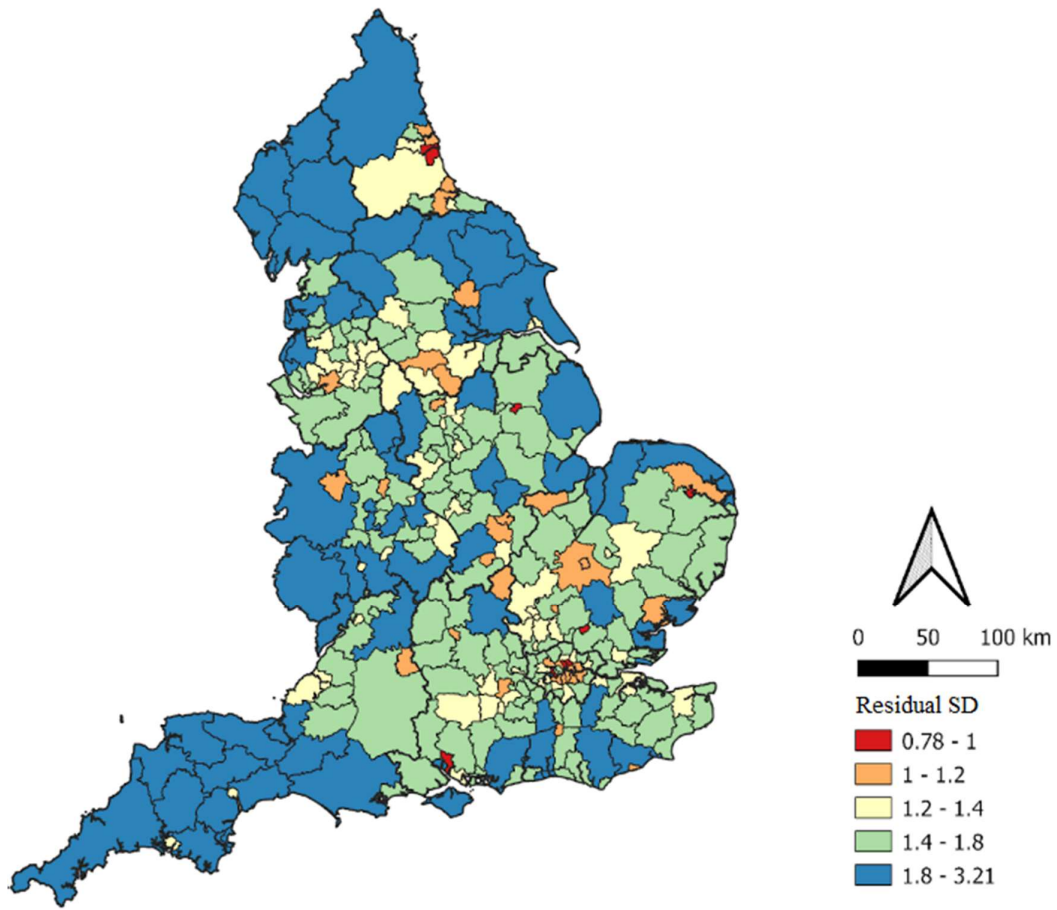

Residuals summaries of the Base Model (including Local Authorities)

|   | *Region* | *SD* | *median* | *range* | *mean* |
|---|---|---|---|---|---|
| 1 | South West | 10.90 | 1.81 | 125 | 0.00 |
| 2 | West Midlands | 10.10 | 1.73 | 118 | 0.00 |
| 3 | North West | 9.80 | 1.56 | 118 | 0.00 |
| 4 | East of England | 9.70 | 1.46 | 116 | 0.00 |
| 5 | East Midlands | 9.70 | 1.51 | 116 | 0.00 |
| 6 | Yorkshire and The Humber | 9.70 | 1.45 | 120 | 0.00 |
| 7 | South East | 9.70 | 1.53 | 120 | 0.00 |
| 8 | London | 9.40 | 1.30 | 113 | 0.00 |
| 9 | North East | 9.10 | 1.34 | 115 | 0.00 |

Table 4  Residual Regression Analysis

| | Specification | adj.R2 | RSE | rsd.median |
|---|---|---|---|---|
| | **Intervention-related Factors** | | | |
| 1 | Fuel | 0.145 | 9.1 | 0.93 |
| 2 | Transaction Type | 0.023 | 9.6 | 1.46 |
| 3 | Glazing | 0.012 | 9.8 | 1.44 |
| 4 | SAP Version + Climate Region | 0.001 | 9.8 | 1.51 |
| | **Location Factors** | | | |
| 5 | Urban/ Rural Divide | 0.0046 | 9.8 | 1.47 |
| 6 | Income Deprivation Percentile | 0.0004 | 9.8 | 1.51 |
| 7 | IMD Percentile | 0.0002 | 9.8 | 1.52 |

Controlling for all immutable characteristics of properties at the LA level tends to overestimate the performance of some homes. Looking at intervention-related factors, a home's primary fuel type captures close to 15% of the prediction error. Relative to homes heated with mains gas in our sample (84%), those with no information provided on fuel type usage (30%) are expected to be almost three rating bands (30pts) off in their estimate. Surprisingly, a property's glazing is less informative than the reason for issuing the EPC certificate (transaction type). Compared to 'sales', most other reasons (e.g. for a survey, with a new build, private and social rent issue) tend to have a slightly larger misestimation (all under 2 pts). Only EPCs issued for assessment (6.5%) – usually for subsidised improvement programmes – tend to have a wider residual range of over 5 pts suggesting a great mix of homes putting themselves forward in the different programmes. The SAP methodology (with its climatic regions provided in the appendix) does not account for much residual variance.

The most important characteristic of location is whether the property is in an urban or rural area. Rural properties are, on average, less efficient than their urban counterparts. Much of this is linked to (lack of) access to mains gas, i.e. fuel type. Rural remoteness could mean access to information, resources, or social pressure might make residents less inclined towards energy improvement, although conversely, if many homes are not already heated by gas central heating, the switch to less polluting and more efficient heating systems such as ground or air-source heat pumps may encounter less resistance if the right incentives are provided. Similarly, as wealth decreases (within each authority) – measured purely in terms of Income Deprivation or as a combination of multiple dimensions – the expected energy efficiency also decreases, indicating that barriers to energy efficiency may be more rooted in affordability than any kind of ideological inertia.

## 3   Discussion and Limitations

Overall, the magnitude of location estimates is small and significant, suggesting that, in theory, locations should not matter that much. However, the poor adj.$R^2$ values - a commonly used metric of model performance -  for both the location and intervention factors suggest more complex interactions might be needed to understand residential energy performance. Energy efficiency scores may be difficult to capture theoretically due to subjective differences in surveyor judgements (DECC, 2014). However, our analysis suggests we need to look closely at the properties which are, on average, mispredicted by ten points or above (approximately one rating band), as these contribute most to low model performance. Our base model (1) is only a starting point that suggests there are a large number of homes we expect should do well that then tend to score much lower. These would be relevant for further investigation and a basis for powerful policy interventions.

**Acknowledgements**

We acknowledge Ordnance Survey for their funding of this project, supported by the Public Sector Geospatial Agreement. This work is part of a larger project with additional team members, including Dr Mark Livingston, Dr Bin Chi and Nick Groom whom the authors are very grateful for their time and feedback.

**Biographies**

Bonnie Buyuklieva is a Research Fellow at the Bartlett Centre for Advanced Spatial Analysis (CASA), where she completed her PhD in 2022. Her research focuses on the intersection of the built environment and population studies through the lens of computational methods. She is interested in themes including sustainability and household health outcomes.

Adam Dennett is a Professor of Urban Analytics and the Head of Department at the Bartlett Centre for Advanced Spatial Analysis (CASA), UCL. Also, the Co-Editor in Chief of Applied Spatial Analysis and Policy, Adam is a Geographer with interests in applied research, broadly in the areas of population, urban inequalities, quantitative methods, GIS, spatial analysis and modelling.

Nick Bailey is Professor of Urban Studies at the University of Glasgow and Director of the Urban Big Data Centre, an international centre-of-excellence in the use of 'digital footprints' data for cities, funded by UKRI-ESRC. His own research focuses mainly on urban housing and neighbourhood systems.

Jeremy Morley has been Chief Geospatial Scientist at Ordnance Survey since 2015. At OS, he leads the Research team, focusing on commissioning, planning and executing research projects with universities, promoting active knowledge transfer and horizon scanning to identify new business opportunities and emerging research.

**Appendix**

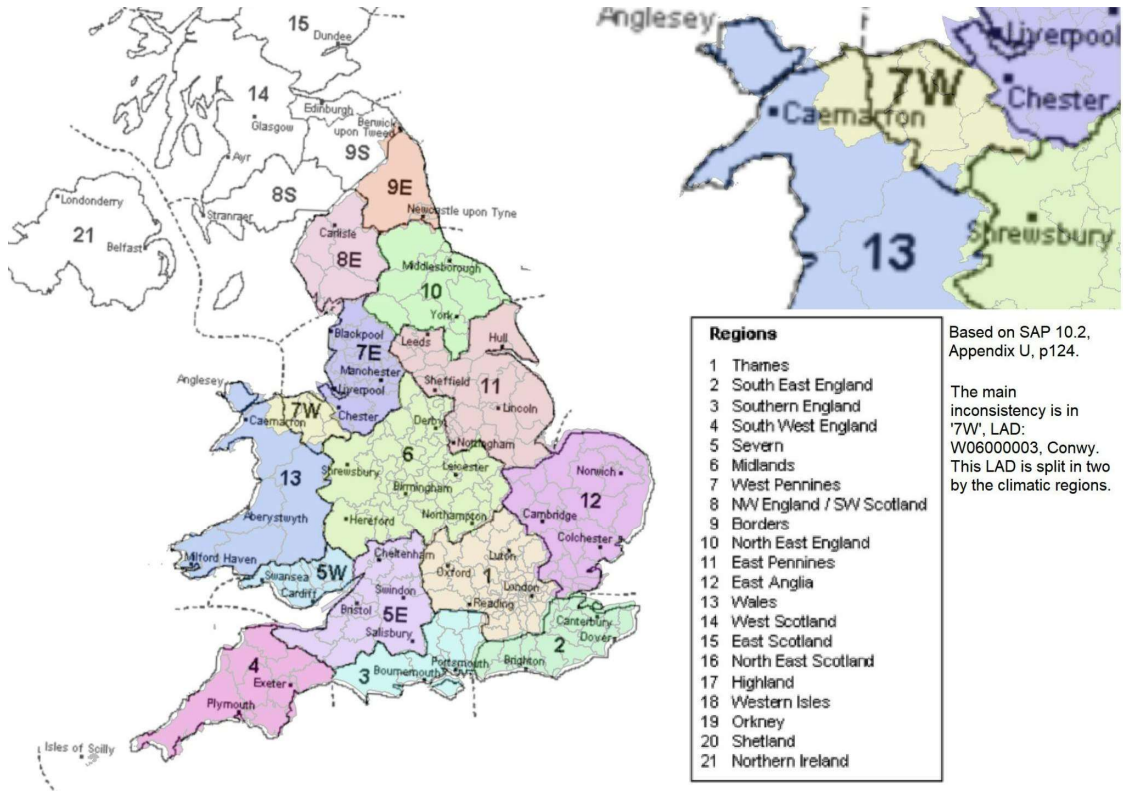

**Figure 2** Climatic Regions from the Standard Assessment Procedure (SAP) Methodology